\documentclass{PoS}
\usepackage{epsfig}

\def\om{\Omega_p}
\def\len{a_B}
\def\lag{D_L}
\def\vpd{{\cal R}}
\def\kin{{\cal V}}
\def\pin{{\cal X}}

\PoS{PoS(BDMH2004)083}

\title{The dark matter content of early-type barred galaxies}

\ShortTitle{Dark matter in early-type barred galaxies}

\author{\speaker{E. M. Corsini}\\
        Dipartimento di Astronomia, Universit\`a di Padova, Italy\\
        E-mail: \email{corsini@pd.astro.it}}

\abstract{
The dynamics of a barred galaxy depends on the pattern speed of its
bar. The only direct method for measuring the pattern speed of a bar
is the Tremaine-Weinberg technique.
This method relies on the analysis of the distribution and dynamics of
the stellar component. It is best suited to gas-poor galaxies and
therefore it has been restricted to early-type barred galaxies.
On the other hand, a variety of indirect methods, which are based on
the analysis of the distribution and dynamics of the gaseous
component, has been used to measure the bar pattern speed in late-type
barred galaxies.
The complete sample of galaxies for which the bar pattern speed has
been directly measured with the Tremaine-Weinberg method is given.
Nearly all the measured bars are as rapidly rotating as they can
be. By comparing this result with recent high-resolution N-body
simulations of bars in cosmologically-motivated dark matter halos, it
is possible to conclude that these bars are not located inside
centrally-concentrated halos.}

\FullConference{Baryons in Dark Matter Halos\\
		 5-9 October 2004\\
		 Novigrad, Croatia}

\begin{document}

\section{Introduction}

Barred galaxies account for roughly half of all disk galaxies in the
local universe [1]. This is also true at higher redshift, because the
rest-frame optical bar fraction is constant out $z\sim1$ [2]. This
makes bars ideal probes of the mass distribution in the central
regions of disk galaxies. Therefore the study of the dynamics of bars
offers an alternative way to constrain the distribution of dark matter
(DM hereafter) in galaxy disks.

The morphology and dynamics of a barred galaxy depend on the pattern
speed of its bar, $\om$. Usually, it is parametrized with the bar
rotation rate $\vpd\equiv\lag/\len$. This is the distance-independent
ratio between the corotation radius, $\lag = V_{\rm c}/\om$, and
the length of bar semi-major axis, $\len$.
At $\lag$ the gravitation and centrifugal forces cancel out in the
rest frame of the bar. $V_{\rm c}$ is the disk circular velocity.
As far as the value of $\vpd$ concerns, if $\vpd < 1.0$ the stellar
orbits are elongated perpendicular to the bar and the bar dissolves.
For this reason, self-consistent bars cannot exist in this regime. Bars
with $\vpd \gtrsim 1.0$ are close to rotate as fast they can, and
there is not a priori reason for $\vpd$ to be significantly larger
than 1.0. Therefore the knowledge of $\vpd$ allows to distinguish
between fast bars ($1.0 \leq \vpd \leq 1.4$) and slow bars ($\vpd >
1.4$). Although there is general consent in setting at $\vpd = 1.4$
the division between fast and slow bars, this choice does not imply
anything about the actual rotation velocity of bar.

\section{Measuring the bar pattern speed}

The only direct method for measuring pattern speeds is the
Tremaine-Weinberg technique [3] (hereafter TW). It gives $\om$ for a
tracer population satisfying the continuity equation. This is the case
of old stellar populations in the absence of significant and patchy
obscuration due to the dust distribution. The TW equation is
$\pin\,\om\,\sin i = \kin$.
where
$\pin = \int\,X\,\Sigma\,dX / \int\,\Sigma\,dX$, 
and 
$\kin = \int\,V_{\rm los}\,\Sigma\,dX / \int\,\Sigma\,dX$
are the luminosity-weighted average of the position $X$ and
line-of-sight velocity $V_{\rm los}$ measured parallel to the major
axis of the galaxy disk, respectively. $\Sigma$ and $i$ are the
surface brightness and disk inclination, respectively.
Slit observations parallel to the major axis of the disk measure all
the quantities needed by the TW equation. 
In fact, for each slit $\pin$ is derived from the surface-brightness
profile, which is obtained by collapsing the galaxy spectrum along the
wavelength direction. On the other hand, $\kin$ is derived from the
one-dimensional spectrum, which is obtained by collapsing the galaxy
spectrum along the spatial direction.
Plotting $\kin$ versus $\pin$ for the different slits produces a
straight line with slope $\om \sin i$.

To date the TW method has been successfully applied to measure the bar
pattern speed of the galaxies listed in Table 1. 
All the galaxies, except for NGC 3992, are SB0's or early-type
spirals. NGC 2950 is the only double-barred galaxy of the sample. For
all the galaxies, $\vpd$ is consistent with being in the range between
1.0 and 1.4. 


This is not a property of early-type barred galaxies only, but it
seems to constitute a generic property of barred galaxies. In fact, a
variety of indirect methods has been used to measure $\om$ and
corresponding $\vpd$ in late-type barred galaxies. They rely on the
identification of morphological features with the location of
Lindblad's resonances, the comparison of the observed gas velocity and
density fields with numerical models of gas flows, and the analysis of
the offset and shape of dust lanes which traces the location shocks in
the gas flows. All these methods are model dependent. Nevertheless,
nearly all the measured bars are found to be fast (see [13] for a
review).
If this result will be confirmed by a successful application of the
TW method to late-type barred galaxies, an important selection bias
present in the current sample of directly measured pattern speeds
would be remedied.
This problem could be addressed by using near-infrared spectroscopy in
order to deal with dust obscuration. [14] applied this technique to
NGC 1068, finding $\vpd<1.0$.  However, they obtained the observables
required by the TW method along two slit positions only, which makes
it hard to constrain $\om$.

\begin{table*}   
\begin{center}   
\caption{Barred galaxies with bar pattern measured by TW method}
\begin{small}
\begin{tabular}{llrrrrrc}   
\noalign{\smallskip}  
\hline 
\noalign{\smallskip}  
\multicolumn{1}{c}{Galaxy}  & 
\multicolumn{1}{c}{Morp. Type} &  
\multicolumn{1}{c}{$D$} &  
\multicolumn{1}{c}{$a_B$}  & 
\multicolumn{1}{c}{$\Omega_B$} & 
\multicolumn{1}{c}{$D_L$} &  
\multicolumn{1}{c}{$\cal{R}$} &
\multicolumn{1}{c}{Ref.} \\
\multicolumn{1}{c}{} & 
\multicolumn{1}{c}{} &
\multicolumn{1}{c}{(Mpc)} &    
\multicolumn{1}{c}{(arcsec)}  & 
\multicolumn{1}{c}{(km s$^{-1}$ arcsec$^{-1}$)} & 
\multicolumn{1}{c}{(arcsec)} &
\multicolumn{1}{c}{} &  
\multicolumn{1}{c}{} \\
\multicolumn{1}{c}{(1)} & 
\multicolumn{1}{c}{(2)} &
\multicolumn{1}{c}{(3)} &    
\multicolumn{1}{c}{(4)} & 
\multicolumn{1}{c}{(5)} & 
\multicolumn{1}{c}{(6)} &
\multicolumn{1}{c}{(7)} &  
\multicolumn{1}{c}{(8)} \\
\noalign{\smallskip}  
\hline 
\noalign{\smallskip}  
ESO 139-G09& (R)SAB0$^0$(rs) & 71.9 & $17^{+6}_{-3}$ & $21.4\pm 5.8$ & $14^{+5}_{-3}$   & $0.8^{+0.3}_{-0.2}$ &  [4]\\ 
ESO 281-G31& SB0$^+$(rs)     & 45.2 & $11\pm1$       & $10.5\pm 4.1$ & $20^{+12}_{-4}$  & $1.8^{+1.1}_{-0.4}$ &  [5]\\ 
IC 874     & SB0$^0$(rs)     & 34.7 & $20\pm5$       & $ 7.0\pm 2.4$ & $27^{+13}_{-7}$  & $1.4^{+0.7}_{-0.4}$ &  [4]\\
NGC 271    & (R')SBab(rs)    & 50.3 & $29\pm1$       & $ 7.8\pm 4.3$ & $44^{+30}_{-16}$ & $1.5^{+1.0}_{-0.5}$ &  [5]\\
NGC 936    & SB0$^+$(rs)     & 14.9 & $50\pm5$       & $ 4.7\pm 1.1$ & $69\pm15$        & $1.4^{+0.5}_{-0.4}$ &  [6]\\
NGC 1023   & SB0$^-$(rs)     &  5.8 & $69\pm5$       & $ 5.1\pm 1.8$ & $53^{+29}_{-14}$ & $0.8^{+0.4}_{-0.2}$ &  [7]\\    
NGC 1308   & SB0/a(r)        & 82.4 & $12^{+2}_{-3}$ & $39.7\pm13.9$ & $9^{+5}_{-2}$    & $0.8^{+0.4}_{-0.2}$ &  [4]\\
NGC 1358   & SAB0/a(r)       & 51.6 & $19\pm3$       & $ 9.3\pm 4.5$ & $23^{+19}_{-7}$  & $1.2^{+1.0}_{-0.4}$ &  [5]\\  
NGC 1440   & (R')SB0$^0$(rs):& 18.4 & $24^{+6}_{-5}$ & $ 7.4\pm 1.7$ & $38^{+11}_{-7}$  & $1.6^{+0.5}_{-0.3}$ &  [4]\\  
NGC 2950   & (R)SB0$^0$(r)   & 19.7 & $34\pm3$       & $11.2\pm 2.4$ & $34^{+9}_{-6}$   & $1.0^{+0.3}_{-0.2}$ &  [8]\\ 
NGC 3412   & SB0$^0$(s)      & 16.0 & $31\pm3$       & $ 4.4\pm 1.2$ & $47^{+17}_{-10}$ & $1.5^{+0.6}_{-0.3}$ &  [4]\\ 
NGC 3992   & SBbc(rs)        & 16.4 & $57\pm12$      & $ 5.7\pm 0.4$ & $45\pm3$         & $0.8\pm0.2$         &  [5]\\ 
NGC 4596   & SB0$^+$(r)      & 29.3 & $66\pm7$       & $ 3.9\pm 1.0$ & $60^{+20}_{-12}$ & $0.9^{+0.5}_{-0.2}$ &  [9]\\ 
NGC 7079   & SB0$^0$(s)      & 34.0 & $25\pm4$       & $ 8.4\pm 0.2$ & $31\pm1$         & $1.2^{+0.3}_{-0.2}$ &  [10]\\ 
\noalign{\smallskip}  
\hline
\noalign{\smallskip}  
\noalign{\smallskip}  
\noalign{\smallskip}  
\end{tabular}
\begin{minipage}{14.8cm}  
NOTE -- 
Col.(2): Morphological classification from RC3, 
         except for ESO 281-G31 (NED). NGC 2950 is a double-barred 
         galaxy and the listed values refer to its primary bar. 
Col.(3): Distance obtained as $V_{{\rm CBR}}/H_0$ with $V_{{\rm CBR}}$ 
         from RC3 and $H_0=75$ km s$^{-1}$ Mpc$^{-1}$.
Col.(4): Bar length. It is from reference papers, except for 
         NGC 936 [11] and NGC 4596 [12].
Col.(5): Bar pattern speed.
Col.(6): Bar corotation radius.
Col.(7): Bar rotation rate.
Col.(8): Reference papers.
\end{minipage}   
\end{small}  
\end{center}   
\end{table*}

\section{Dark matter distribution in barred galaxies} 

Theory and N-body simulations favor the birth of fast bars, while the
time evolution of $\vpd$ depends on the DM distribution of the galaxy.
Direct measurement of bar pattern speeds are consistent with bars
being fast rotators. This conclusion places important constraints on
the relative fraction of luminous to dark matter at least inside the
optical region of early-type barred galaxies.

The dynamical friction with a dense DM halo brakes the bar on a short
time scale compared with the ages of galaxies [15]. As a bar slows
down, $\vpd$ increases to values larger than 1.4. Fast bars require
that the disk, in which they formed, contributes most of the
rotational support in the inner parts of the galaxy [16]. This means
that barred galaxies have a maximal disk, instead of a
centrally-concentrated DM halo.
This conclusion holds also for bright unbarred galaxies. They have
comparable fractions of DM at a given radius as do their barred
counterparts, as it results from both the comparison of their
Tully-Fisher relations [16,17] and the analysis of their mass
distribution [e.g., 18].

At present there is a lively debate about this suggestion for barred
galaxies. For example, some high-resolution N-body simulations with
centrally-concentrated DM halos produce bars with $\vpd=1.7$ [19]. It
has been argued that these slow bars are still consistent with data
given in Table 1. However, it should be noticed that this limit is
reached only in those galaxies with the largest uncertainties and
ignores other source of scatter [20]. Nevertheless, it is clear that
there is still the need of an accurate measurement of $\vpd$ in a
statistically significant number of barred galaxies. Direct
measurements of $\om$ with the TW method using near-infrared
spectroscopy will open the possibility of a systematical investigation
of $\vpd$ not only in early-type but also in late-type barred
galaxies.

\bigskip
\noindent
{\bf Acknowledgments.} It is a pleasure to thank V.P. Debattista and
J.A.L. Aguerri for helpful discussions and collaborate with them in
measuring bar pattern speeds.

\end{document}